\documentclass[conference]{IEEEtran}
\IEEEoverridecommandlockouts
\usepackage{cite}
\usepackage{amsmath,amssymb,amsfonts}
\usepackage{algorithmic}
\usepackage{graphicx}
\usepackage{textcomp}
\usepackage{xcolor}
\usepackage{hyperref}
\usepackage{multirow}
\usepackage{booktabs}
\usepackage{diagbox}
\def\BibTeX{{\rm B\kern-.05em{\sc i\kern-.025em b}\kern-.08em
    T\kern-.1667em\lower.7ex\hbox{E}\kern-.125emX}}
\begin{document}

\title{Uncovering the Visual Contribution in Audio-Visual Speech Recognition

\thanks{This work was conducted with the financial support of the Research Ireland Centre for Research Training in Digitally-Enhanced Reality (d-real) under Grant No. 18/CRT/6224.}
}

\author{
\IEEEauthorblockN{Zhaofeng Lin, Naomi Harte} 
\IEEEauthorblockA{\textit{Sigmedia Group, School of Engineering} \\
\textit{Trinity College Dublin}\\
Ireland \\
\{linzh, nharte\}@tcd.ie}
}

\maketitle

\begin{abstract}
Audio-Visual Speech Recognition (AVSR) combines auditory and visual speech cues to enhance the accuracy and robustness of speech recognition systems. 
Recent advancements in AVSR have improved performance in noisy environments compared to audio-only counterparts. 
However, the true extent of the visual contribution,  and whether AVSR systems fully exploit the available cues in the visual domain, remains unclear. 
This paper assesses AVSR systems from a different perspective, by considering human speech perception. 
We use three systems: Auto-AVSR, AVEC and AV-RelScore. 
We first quantify the visual contribution using effective SNR gains at 0 dB and then investigate the use of visual information in terms of its temporal distribution and word-level informativeness. 
We show that low WER does not guarantee high SNR gains. 
Our results suggest that current methods do not fully exploit visual information, and we recommend future research to report effective SNR gains alongside WERs.

\end{abstract}

\begin{IEEEkeywords}
Audio-visual speech recognition, Visual contribution
\end{IEEEkeywords}

\section{Introduction}
\label{sec:intro}
Humans perceive speech in a multimodal manner; we watch and listen when we communicate with people \cite{mcgurk1976hearing}. 
Viewing a speaker's facial and lip movements enhances speech intelligibility and the contribution of visual information to speech perception becomes even more significant in challenging environments, such as those with background noise \cite{sumby1954visual, macleod1987quantifying}.

Audio-Visual Speech Recognition (AVSR) systems exploit visual information from the speaker’s face, e.g. lip movements, to enhance speech recognition. This area has attracted significant research attention in recent years. The rapid development of artificial intelligence and end-to-end modelling for Automatic Speech Recognition \cite{li2022recent} has led to significant progress in AVSR, resulting in improved performance compared to audio-only systems across various acoustic environments \cite{afouras2018deep, sterpu2018attention, petridis2018end}.

Despite these advancements, the development of AVSR seems to have reached a ``bottleneck" stage. State-of-the-art (SOTA) models continue to demonstrate improved performance, i.e. lower word error rates (WERs) and greater robustness, compared to their audio-only counterparts in noise.
However, much of the current research remains focused on metric-driven improvements, i.e. incremental reductions to WER, with less attention given to understanding how visual information enhances these systems or whether the visual component is being fully exploited in existing approaches.

This paper takes a step back to assess SOTA systems from a different perspective by considering human speech perception. Through this, we hope to gain insights as to whether the visual component is being fully exploited in existing AVSR systems, helping to uncover better approaches.
Inspired by human speech perception studies, we explore this with three questions: 
1) Can metrics beyond WER better reflect visual contributions?
2) How does the presence and timing of visual information influence AVSR systems?
3) Do AVSR systems perform better for words with higher levels of visual informativeness, as determined by human speech perception?

Using three SOTA AVSR systems, we revisit effective SNR gains  \cite{potamianos2001large} to evaluate visual gains. We also present two experiments. The first occludes the mouth region at different stages of words 
to assess the impact of lip movements. A second experiment compares errors made by audio-only, video-only, and AVSR systems with those made in human audiovisual speech perception.  The key contributions of this paper are:
\begin{itemize}
   \item We quantify the visual contribution of AVSR systems using effective SNR gains.
   \item We reveal a discrepancy in how AVSR systems use temporal visual information compared to human reliance on early visual cues.
   \item Our empirical investigation of the relation between AVSR errors and visual informativeness for words.
\end{itemize}

\section{Related Research}
\label{sec:relatedWorks}
\subsection{Audio-visual Speech Perception}

In this section, we explore established contributions of visual information to speech perception, examining how temporal distribution and visual informativeness have been studied in both human perception and AVSR systems. These insights motivate our alternate approach to assessing  AVSR systems.

\subsubsection{Quantifying Visual Contribution}
Many studies have attempted to quantify the contribution of visual information to speech perception \cite{macleod1987quantifying, erber1969interaction, potamianos2001large}.
MacLeod and Summerfield \cite{macleod1987quantifying} measured auditory and audio-visual speech-reception thresholds (SRTs) in human listeners. The SRT represents the signal-to-noise ratio (SNR) in dB at which participants correctly recognise all three keywords in a speech task. The visual benefit was quantified as the difference in dB between the SRTs obtained in auditory-only and audio-visual conditions.
In a 2001 study, Potamianos et al. \cite{potamianos2001large} compared AVSR systems with human perception of audio-visual speech. They quantified the visual benefit in AVSR as an effective SNR gain, measured by the difference in SNR at which the AVSR WER equals the reference WER for audio-only recognition at 10 dB.
They observed approximately a 6 dB effective SNR gain compared to the audio-only performance at 10 dB for both machines and humans. 
These findings provide a useful benchmark for our work to evaluate how effectively current AVSR systems are leveraging visual information compared to human performance.

\subsubsection{Temporal Distribution of Information}
The distribution of information in audio-visual speech varies temporally across phonemes and words.
Jesse and Massaro \cite{jesse2010temporal} found that the timing of information availability differs, with visual speech cues often fully accessible earlier in the phoneme and words, while auditory information accumulates over time.
This early visual information, although supplementary to auditory cues, provides significant early recognition benefits, particularly for visually defined linguistic features like the place of articulation.
Consequently, visual speech contributes more to recognition at the beginning of a word than toward its end. 
Karas et al. \cite{karas2019visual} further demonstrated that visual speech can suppress auditory responses, especially for words with a visual head start. 
For example, in an audiovisual recording of the word ‘drive' the visual onset of the open mouth for the initial phoneme ‘d' precedes auditory vocalisation by 400 ms, indicating an auditory-visual asynchrony. 
These findings highlight the importance of timing in visual speech perception and the difference in visual information within a word. 
In this paper, we investigate the impact of the absence of lip movement at different stages of speech on AVSR, taking into account the early visual benefits.

\subsubsection{MaFI Score Overview}
Krason et al. \cite{krason2023mouth} introduced the concept of Mouth and Facial Informativeness (MaFI) for words, quantifying visual informativeness based on mouth and facial movements. MaFI scores are derived from a speechreading task where participants watch silent video clips and guess the words being spoken.
The process of computing MaFI scores involves three main steps. First, both the target words and participants' guesses are converted into International Phonetic Alphabet (IPA) segments. Next, these IPA segments are transformed into phonological feature vectors. Finally, the similarity between the phonological features of the IPA segments is calculated as MaFI scores. The resulting scores range from approximately -2.5 to 0, with higher scores (closer to 0) indicating greater visual informativeness.
MaFI scores were shown to capture the dynamic nature of mouth and facial movements well \cite{krason2023mouth}, with words containing phonemes featuring roundness, frontness, or visemes like lower lip tuck, lip rounding, and lip closure being more visually informative.
In this paper, we use MaFI scores to assess how well current AVSR systems recognise visually informative words.

\subsection{AVSR Systems}
In this paper, we used three open-source, trained, SOTA AVSR systems: Auto-AVSR \cite{ma2023auto}\footnote{\url{https://github.com/mpc001/Visual_Speech_Recognition_for_Multiple_Languages}}, 
Audio-Visual Efficient Conformer (AVEC) \cite{burchi2023audio}\footnote{\url{https://github.com/burchim/AVEC}} 
and AV-RelScore \cite{hong2023watch}\footnote{\url{https://github.com/ms-dot-k/AVSR}}. 
All systems were published in 2023 and take raw audio waveforms as input, with visual inputs consisting of cropped mouth regions at 96 × 96 pixels. Their performance across different noise levels on LRS2 \cite{son2017lip} and LRS3 \cite{afouras2018lrs3} is illustrated in Fig. \ref{fig:snr_gains}

\subsubsection{Auto-AVSR}
This system is built on a CTC/Attention model based on
a ResNet-18 and Conformer architecture \cite{ma2021end}.
The audio front-end module is based on a ResNet-18 \cite{he2016deep} architecture with 1D convolutional layers, while the visual front-end module uses a modified ResNet-18, where the first convolutional layer has been replaced by a 3D convolutional layer.
Both audio and visual features are then encoded using a Conformer encoder \cite{gulati20_interspeech}, which serves as the back-end for temporal modelling. 
For modality fusion, the encoded acoustic and visual features from back-end modules are concatenated and projected to the same dimensional space as the visual features using a multi-layer perceptron (MLP).
The rest of the network includes a projection layer and a Transformer decoder for joint CTC/attention training \cite{watanabe2017hybrid}.
Auto-AVSR was trained on both labelled corpora (LRW \cite{chung2017lip}, LRS2, and LRS3) and unlabelled corpora (AVSpeech \cite{47257} and VoxCeleb2 \cite{chung18b_interspeech}), in total 3,448 hours of data.

\subsubsection{AVEC}
This model evolved from the Efficient Conformer architecture \cite{burchi2021efficient}, used as the back-end network for processing both audio and visual modalities. The Efficient Conformer encoder consists of multiple stages, with each stage comprising several Conformer blocks. 
Additionally, intermediate CTC losses were applied between Conformer encoders. This approach was claimed to enhance visual speech recognition performance and address a common issue where audio-visual models tend to ignore the visual modality.
AVEC is a non-autoregressive method with CTC loss only. It was trained on the LRW, LRS2 and LRS3 datasets, in total 818 hours of data.

\subsubsection{AV-RelScore}
This model adopted much of the architecture from Auto-AVSR, with contributions in two key areas. First, it introduced audio-visual corruption modelling, including lip occlusions, which greatly enhanced robustness. Second, it proposed an Audio-Visual Reliability Scoring (AV-RelScore) module, designed to weigh input modalities according to their reliability.
It is important to note that, unlike the other two systems that trained models using combined datasets, this study trained models separately on LRS2 and LRS3.

\section{Datasets}
All experiments in this paper used LRS2-BBC \cite{son2017lip} and LRS3-TED datasets \cite{afouras2018lrs3}.
The LRS2-BBC dataset, known as LRS2, includes 224.1 hours of video containing 144,482 clips from BBC television. This dataset is divided into several subsets: 96,318 utterances for pre-training (195 hours), 45,839 for training (28 hours), 1,082 for validation (0.6 hours), and 1,243 for testing (0.5 hours). The video files in LRS2 have a resolution of 160 × 160 pixels and a frame rate of 25 fps.

The LRS3-TED dataset, known as LRS3, comprises over 400 hours of video sourced from 5594 TED and TEDx talks in English, downloaded from YouTube. LRS3 contains 151,819 utterances (439 hours). The breakdown of this dataset includes 118,516 utterances in the pretraining set (408 hours), 31,982 in the training-validation set (30 hours), and 1,321 in the test set (0.9 hours). The video files in the dataset have a resolution of 224 × 224 pixels and a frame rate of 25 fps.

\section{Effective SNR Gains}
The effective SNR gain, as proposed in \cite{potamianos2001large}, is measured by the difference in SNR at which the AVSR WER equals the reference WER for audio-only recognition at 10 dB. 
This metric quantifies the benefit of the visual modality in reducing WER compared to audio-only speech recognition across various acoustic channel SNRs.
However, with the development of modern neural architectures, systems generally perform well at 10 dB, making SNR gains at this level less effective for assessing performance in noisy environments. Therefore, this paper proposes quantifying SNR gains with reference to the audio-only WER at 0 dB.

\subsection{Results}
\begin{figure}[t]
  \centering
  \includegraphics[width=0.7\linewidth]{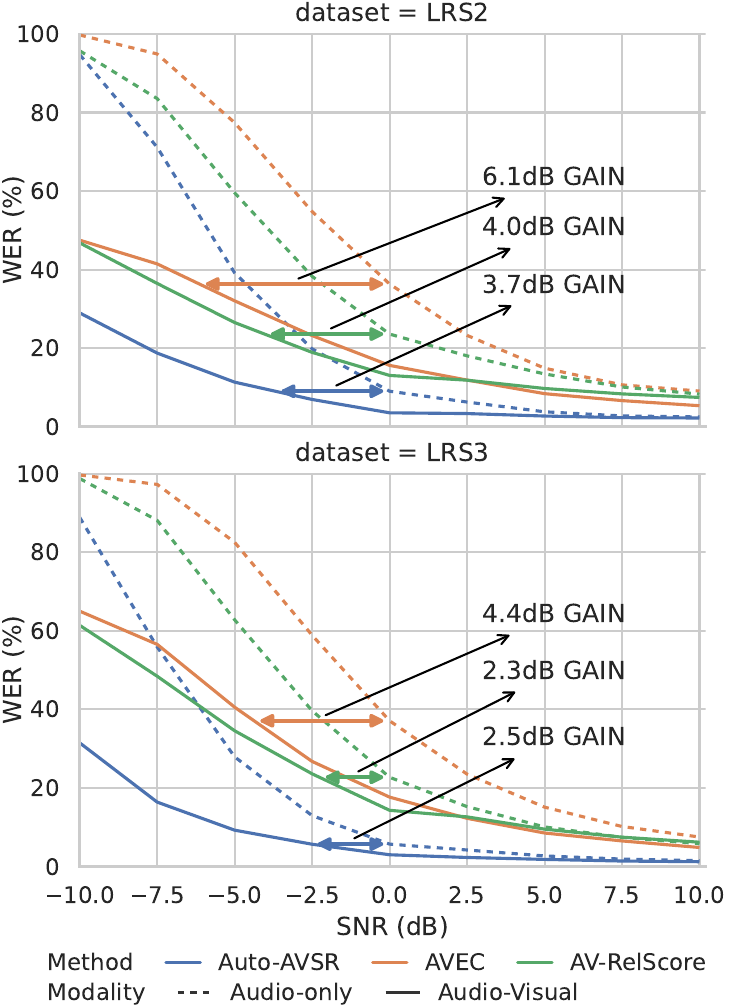}
  \caption{Performance of six models (three methods, both audio-only and audio-visual) on LRS2 and LRS3 test sets under different levels of pink noise.}
  \label{fig:snr_gains}
\end{figure}

Figure \ref{fig:snr_gains} shows the audio-only and audio-visual performance of three methods across various SNR conditions on LRS2 and LRS3 test sets. 
The effective SNR gains were measured compared to audio-only performance at 0 dB. 

Auto-AVSR performs best across all SNR conditions, likely due to its notably larger training dataset. 
However, it has the lowest effective SNR gain of 3.7 dB on LRS2 and 2.5 dB on LRS3.
AV-RelScore, performing worse than Auto-AVSR overall in WER, achieves a higher effective SNR gain of 4.0 dB on LRS2 and the lowest of 2.3 dB on LRS3. 
AVEC, despite performing the worst in noisy conditions, has the highest effective SNR gains, at 6.1 dB on LRS2 and 4.4 dB on LRS3.

As reported in \cite{potamianos2001large}, both machines and humans achieved an effective SNR gain of 6 dB when compared to audio-only performance at 10 dB on the IBM ViaVoice audio-visual database. 
Although all three methods here were tested on LRS2 and LRS3, none reached this level of effective SNR gain when compared at 10 dB. 

Interestingly, Auto-AVSR performing best in WER has the lowest effective SNR gain, while AVEC, with the worst WER, has the highest SNR gain. This indicates that top-performing systems in terms of WER, may not fully leverage visual information. 
It highlights the limitation of only reporting WERs to evaluate AVSR performance.

\section{Occlusion Tests}
This experiment assesses how current AVSR systems utilise different temporal segments of visual information, specifically from the mouth region, given that visual speech information is fully available earlier in words. 
Therefore, we hypothesize that AVSR systems will perform worse with initial occlusion than with middle occlusion.

\subsection{Setup}
\begin{figure}[tbh]
  \centering
  \includegraphics[width=0.8\linewidth]{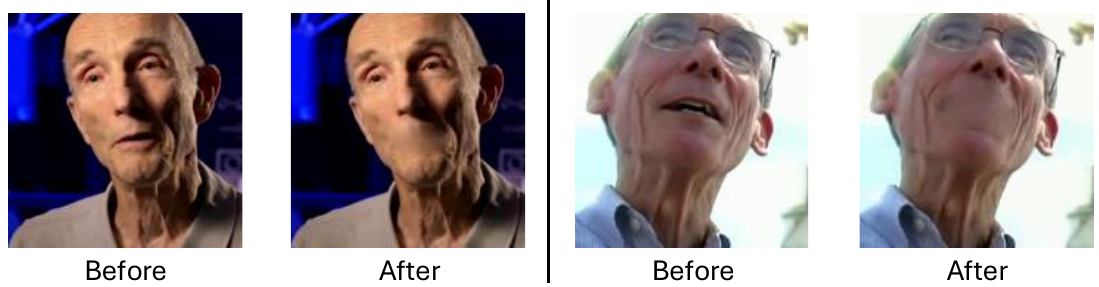}
  \caption{Examples of video frames before and after inpainting occlusion.}
  \label{fig:occlude_exp}
\end{figure}

To occlude the video, we first used Montreal Forced Aligner \cite{mcauliffe2017montreal} to extract timestamps for the start and end of every word in all utterances within the datasets. These timestamps, initially recorded in seconds, were then mapped to corresponding video frames to identify the specific frames that marked the beginning and end of each word.
Next, we applied image inpainting techniques to selectively occlude the lip region in these frames, generating video frames where the speaker's lips were no longer visible. Specifically, we separately occluded the initial and middle 1/3 of the frames for each word in the video, resulting in 2 different test sets. For words lasting more than two frames, 1/3 of their frames were occluded; for words lasting fewer than three frames, no occlusion was applied to avoid extreme distortion.
Figure \ref{fig:occlude_exp} illustrates examples of video frames from LRS2 before and after inpainting occlusion, where the lip region is no longer visible.

In this experiment, Auto-AVSR, AVEC, and AV-RelScore were tested on the LRS2 test set under three conditions: without occlusion, with initial occlusion, and with middle occlusion, all within a pink noise environment at 0 dB SNR.

\subsection{Results}
Table \ref{tab:exp1} shows the WERs of each AVSR model on LRS2 test set with no occlusion, initial occlusion, and middle occlusion in pink noise at 0 dB. 
Results on LRS3 were similar to those on LRS2 and are omitted here due to space constraints.

Auto-AVSR suffers from both initial and middle occlusions, with a notable relative increase in WER of around 70\% for both types of occlusions.
AVEC, conversely, is notably more reliant on the middle segment of visual information.
The initial occlusion caused a relative 58\% WER increase and 73\% caused by middle occlusion. 
AV-RelScore shows a very consistent reliance on both initial and middle segments of visual information. It exhibits a relative increase in WER, with a 65\% relative increase for both initial and middle occlusions.

Overall, Auto-AVSR and AV-RelScore exhibit a similar pattern, both suffering from high increases in WER under occlusions. 
In contrast, AVEC is more affected by middle occlusion than initial occlusion, indicating a greater reliance on the middle segment. 
These results show varying patterns in how AVSR systems use temporal visual information: some maintain a consistent reliance across segments, while others focus more on information in the middle.
Thus, there is a mismatch between AVSR systems and human speech perception, as humans benefit more from visual cues at the beginning of words, whereas AVSR systems show different dependencies.

\begin{table}[t]
\centering
\caption{WERs of three AVSR models on LRS2 with no occlusion, initial occlusion, and middle occlusion in pink noise at 0 dB.}
\begin{tabular}{l|ccc} 
\toprule
\textbf{Model}       & \textbf{None} & \textbf{Initial} & \textbf{Middle}  \\ 
\midrule
\textbf{Auto-AVSR}   & 3.5          & 6.1          & 5.9            \\
\textbf{AVEC}        & 15.6         & 24.6            & 27.0           \\
\textbf{AV-RelScore} & 13.0         & 21.5           & 21.5            \\
\bottomrule
\end{tabular}
\label{tab:exp1}
\end{table}

\section{MaFI Score Analysis}

This experiment explores the relationship between AVSR system errors and visual word informativeness, as measured by the MaFI score. Specifically, we determine whether words with higher MaFI scores correlate with fewer recognition errors.

\subsection{Setup}
To investigate this relationship, we calculated the Individual Word Error Rate (IWER) \cite{goldwater2010words} for each word. The IWER is calculated as the sum of substitution and deletion errors divided by the total occurrences of the word, allowing us to quantify the error rate at the word level. While substitution and deletion errors can be directly aligned with reference words, insertion errors are excluded from this analysis due to the difficulty in attributing them to specific words.
We conducted Pearson’s correlation analysis between MaFI scores and IWERs for all audio-only, video-only, and audio-visual models of Auto-AVSR and AVEC. 

It is important to note that, out of the 2,276 words with MaFI scores, the LRS2 test set contains 1,697 words, with only 593 words overlapping between the two datasets. Additionally, IWERs tend to be either 1 or 0 for low-frequency words, so our analysis focuses on words that occur more than six times.

\subsection{Results}
Table \ref{tab:exp2} presents Pearson’s correlation analysis between MaFI scores and IWERs for audio-only (AO), video-only (VO), and audio-visual (AV) models of Auto-AVSR and AVEC (AV-RelScore did not provide their VO models) under noisy conditions at -5, 0, and 5 dB.

In AO mode, no significant correlation between MaFI scores and IWERs is observed in either model, confirming that visual informativeness does not influence audio-only recognition. 
In contrast, both Auto-AVSR and AVEC show a slight yet statistically significant negative correlation in VO mode (p-value $<$ 0.01), indicating that higher visual informativeness is associated with fewer recognition errors when relying solely on visual cues. 
However, in AV mode, the models diverge: Auto-AVSR displays a weaker, insignificant correlation, suggesting reduced reliance on visual cues when audio is present, while AVEC maintains a significant negative correlation, particularly at lower SNR levels (-5 and 0 dB), demonstrating effective use of visual information in noisy environments.

\begin{table}[t]
\centering
\caption{Pearson’s correlation test between MaFI scores and IWERs. ** ($p < 0.01$); ***($p < 0.001$). }
\setlength{\tabcolsep}{0.7\tabcolsep}
\begin{tabular}{c|ccc||ccc} 
\toprule
\textbf{Model}       & \multicolumn{3}{c||}{\textbf{Auto-AVSR}}             & \multicolumn{3}{c}{\textbf{AVEC}}                        \\ 
\midrule
\textbf{SNR}         & \textbf{-5} & \textbf{0}                & \textbf{5} & \textbf{-5} & \textbf{0}                   & \textbf{5}  \\ 
\midrule
\textbf{\textbf{AO}} & 0.002       & -0.021                    & 0.039      & -0.019      & -0.020                       & -0.052     \\
\textbf{VO}          & \multicolumn{3}{c||}{-0.097**}                       & \multicolumn{3}{c}{-0.173***}                            \\
\textbf{\textbf{AV}} & -0.040      & \multicolumn{1}{l}{0.055} & 0.035      & -0.151**     & \multicolumn{1}{l}{-0.150**} & -0.080      \\
\bottomrule
\end{tabular}
\label{tab:exp2}
\end{table}

\section{Discussion and Conclusion}
This paper investigates how AVSR systems exploit visual information. 
By quantifying effective SNR gains for SOTA systems at 0 dB, we observed that while the system with the best WER had the lowest effective SNR gain, the worst-performing system had the highest gain.
Occlusion tests revealed different patterns in using temporal visual information, and a mismatch with human speech perception, as AVSR systems do not benefit more from early visual cues like humans.
MaFI score analysis demonstrated that AVEC effectively uses visual information, especially in noisy conditions, whereas Auto-AVSR's weaker correlation suggests a greater reliance on audio. 
Overall, we observe that AVEC exploits visual information better in both effective SNR gains and MaFI experiments. 
This may be attributable to the intermediate CTC, which forces pre-fusion layers to learn spatiotemporal features.

We believe that the current SOTA methods can do more to fully exploit visual information, and that human speech perception abilities should set a target for our expectations of what visual cues can offer speech recognition in noise.
We also recommend researchers report effective SNR gains alongside WERs in their papers, to provide a more comprehensive assessment of AVSR performance.

\vfill\pagebreak

\bibliographystyle{IEEEtran}
\bibliography{refs}

\end{document}